\documentclass[%
 aip,
 amsmath,amssymb,
 reprint,%
]{revtex4-1}

\usepackage{graphicx}
\usepackage{dcolumn}
\usepackage{bm}

\usepackage[utf8]{inputenc}
\usepackage[T1]{fontenc}
\usepackage{mathptmx}
\usepackage{etoolbox}

\makeatletter
\def\@email#1#2{%
 \endgroup
 \patchcmd{\titleblock@produce}
  {\frontmatter@RRAPformat}
  {\frontmatter@RRAPformat{\produce@RRAP{*#1\href{mailto:#2}{#2}}}\frontmatter@RRAPformat}
  {}{}
}%
\makeatother
\begin{document}

\preprint{AIP/123-QED}

\title{Cascading failure with memory effect in random networks}
\author{Yanpeng Zhu}
\author{Lei Chen}%
\author{Fanyuan Meng}
\author{Chun-Xiao Jia}
\author{Run-Ran Liu}
\email{fanyuan.meng@hznu.edu.cn;chuanxiaojia@163.com;runranliu@163.com}
\affiliation{ 
Research Center for Complexity Sciences, Hangzhou Normal University
}%

\date{\today}

\begin{abstract}
In many cases of attacks or failures, memory effects play a significant role. Therefore, we present a model that not only considers the dependencies between nodes but also incorporates the memory effects of attacks. Our research demonstrates that the survival probability of a random node reached by a random edge surpasses the inverse of the average degree ($1/{\langle k \rangle}$), and a giant component emerges regardless of the strength of dependencies. Moreover, if the dependency strength exceeds $1/{\langle k \rangle}$, the network experiences an abrupt collapse when an infinitesimally small fraction of nodes is removed, irrespective of the memory effect. Our proposed model provides insights into the interplay between dependencies between nodes, memory effects, and the network structures under attacks or failures. By considering these factors, we can better assess the vulnerability of complex systems and develop strategies to mitigate cascading failures.
\end{abstract}

\maketitle

\section{Introduction}

Cascading failures, characterized by the propagation of failures across interconnected components in a system, have garnered significant interest and concern across various domains \cite{valdez2020cascading,duenas2009cascading,crucitti2004model}. This phenomenon has particularly raised attention in fields such as power grids, transportation networks, financial systems, and social networks. For instance, in power grids, the failure of a single transmission line or power plant can generate an overload on the remaining infrastructure, triggering further failures and potentially leading to widespread blackouts \cite{schafer2018dynamically,pahwa2014abruptness}. Similarly, in transportation networks, a disruption such as a traffic accident or severe weather conditions can result in congestion and delays, affecting the entire network and initiating cascading failures \cite{candelieri2019vulnerability,li2022exploring, wu2022cascading}. Financial systems are also susceptible to cascading failures, whereby the failure of a large bank or financial institution can trigger a loss of confidence, subsequently causing failures in other interconnected entities and potentially culminating in a financial crisis \cite{ramirez2023stochastic, huang2013cascading}. Furthermore, social networks can exhibit cascading failures wherein the spread of misinformation or viral content can fuel the amplification of negative behaviors or undermine trust and cooperation, ultimately resulting in the collapse of social dynamics  \cite{yi2015modeling}. The consequences of cascading failures are quite severe as described above, encompassing widespread disruptions, system-wide collapses, significant economic losses, and potential threats to human safety and well-being.

Cascading failures have been extensively studied using various models that explore different potential sources for their origin. These models aim to investigate the underlying mechanisms and dynamics of cascading failures. One such potential source is the overload of a node or link due to load redistribution \cite{motter2002cascade,baxter2010bootstrap,holme2002vertex,holme2002edge}. When the load on a specific node or link increases beyond its capacity, it can lead to its failure, subsequently triggering a cascade of failures in interconnected components. Another source of cascading failures is the presence of direct dependencies among nodes \cite{parshani2011critical, buldyrev2010catastrophic, liu2016cascading, liu2018weak}. In systems where nodes depend on one another if a particular node fails, it can cause the failure of all nodes that depend on it. This dependency-driven cascade can propagate through the system, causing widespread failures. An additional source is that the number of multifunctional neighbors surrounding a node exceeds a threshold value, like the global cascades model, bootstrap percolation, and $k$-core percolation \cite{watts2002simple, cellai2011tricritical, dorogovtsev2006k}. Here, researchers often assume the Markovian nature of the system. This assumption means that the probability of a node or link failure depends solely on the current state of the system, disregarding its past states or events (without memory). This simplification enables the application of well-established mathematical techniques and facilitates the analysis and prediction of cascading failures.

However, real-world systems often exhibit non-Markovian behavior. This non-Markovian property is observed in various domains, including epidemic spreading processes \cite{feng2019equivalence, starnini2017equivalence, tomovski2021discrete}, diffusion processes \cite{lenzi2010non,mura2008non}, information spreading \cite{gao2017promoting,xue2013non}, and synchronization \cite{casado2005theory} which could be strongly related to the cascading process. While the Markovian assumption simplifies the dynamics of complex systems, it overlooks the fact that the failure probability of a node is influenced not only by its current state but also by its historical sequence of events leading up to the present. For example, where the recovery process from failures depends on past states and events, making them more resilient or vulnerable against failures \cite{lin2020non}. 

Thus it is crucial to introduce non-Markovian properties into current research to gain a deeper understanding of cascading failures. To this end, we propose a model to mimic the cascading failure process by introducing both node dependency and non-Markovian effects of attacks. We assume each pair of nodes has a dependency strength, i.e. one node's failure would trigger each neighborhood node with a certain probability, but the waves of attack on every node have a memory effect. We find that the interplay between dependency strength, memory effects, and network structure plays a crucial role in determining the system's phase transition types, and critical points, as well as its robustness. Our research indicates that when the survival probability of a random node reached by a random edge surpasses the inverse of the average degree ($1/{\langle k \rangle}$), a giant component emerges regardless of the strength of dependencies. Moreover, if the dependency strength exceeds $1/{\langle k \rangle}$, the network experiences an abrupt collapse when an infinitesimally small fraction of nodes is removed, irrespective of the memory effect. Additionally, positive or negative excessive memory effects of attacks can render the system more fragile or robust. 

\section{Model}
A random network with $N$ nodes is constructed in which the degree $k$ of each node follows the Poisson distribution $p(k)=\frac{e^{-\langle k \rangle}{\langle k \rangle}^k}{k!}$ and mean degree is $\langle k \rangle$. Initially, a fraction  $1-p$ of nodes is removed from the network, where $p$ lies between 0 and 1. We assume that each pair of connected nodes has a basic dependence strength of $\alpha \in [0,1]$, indicating the failure of one node can trigger the failure of the other node with probability $\alpha$. Importantly, a node's fragility can exhibit non-Markovian behavior, which implies a memory effect. This means that the fragility of a node can be influenced not only by the current state but also by the past history of attacks. Specifically, if a node successfully withstands an attack from a dependent neighborhood node's failure, its fragility may either become stronger or weaker, depending on the situation. To capture both effects on the cascading failure process, we introduce a function denoted as $f(\alpha,b,n)$ in our model. Here, $n$ represents the number of attacks, and $b$ denotes the degree of impact accumulated after each failed attack. A positive $b$ indicates an increasing impact on the fragility of each node, while a negative $b$ implies a decreasing impact. The function $f(\alpha,b,n)$ is formulated as follows
\begin{equation} \label{eq:tanh}
   f(\alpha,b,n) =
\begin{cases}
\alpha + (1-\alpha) \tanh(b(n-1))       & b \geq 0,n \neq 0, \\
\alpha + \alpha \tanh (b(n-1))           & b \leq 0,n \neq 0, \\
0                                       & n = 0,
\end{cases}
\end{equation}
where $\tanh$ is the hyperbolic tangent function. In the case of $b \geq 0$ and $n \neq 0$, the hyperbolic tangent function is employed to progressively enhance the dependence strength from $\alpha$ to $1$. This adjustment signifies an increase in the node's fragility after surviving previous attacks. Conversely, the hyperbolic tangent function gradually adjusts the dependence strength from $\alpha$ to $0$, indicating a reduction in the node's fragility after surviving previous attacks. Lastly, when no attacks have occurred ($n = 0$), the function sets the dependence strength to $0$, implying no influence from failed neighbors. The process of cascading failures comes to a halt when no further nodes experience failure and the system stabilizes.

\begin{figure}[htb]
    \centering
    \includegraphics[width=\linewidth]{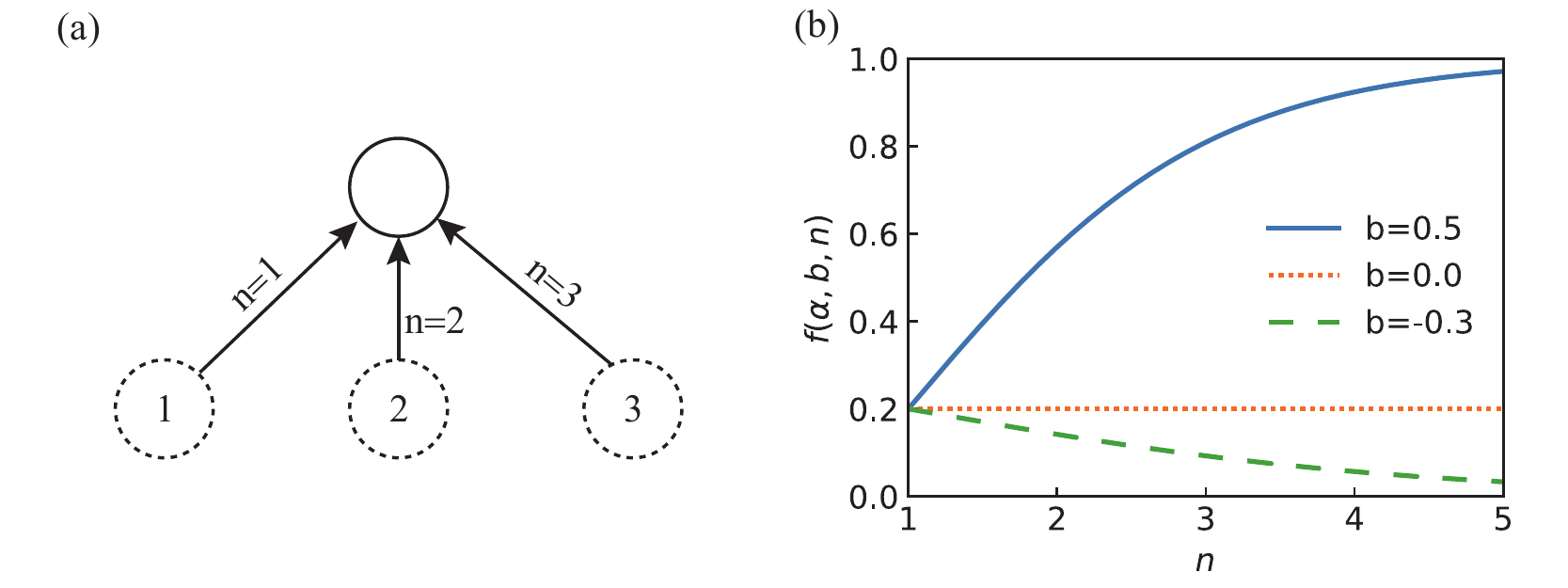}
    \caption{(a) A schematic diagram illustrates the non-Markovian impact of the failed neighbors on the central node; (b) The illustration of the function $f(\alpha,b,n)$ with $\alpha = 0.2$ for different $b$ values.}
    \label{fig:fig1}
\end{figure}

Fig.\ref{fig:fig1}(a) provides a visualization of the impact of failed neighbors on the central node in our model. Let's consider the sequential failure of nodes 1, 2, and 3 in a step-by-step manner. Initially, when node 1 fails, it triggers the central node to potentially fail with a probability $\alpha$, as determined by Eq.\ref{eq:tanh}. If the central node manages to survive, the presence of a memory effect or non-Markovian node dependency becomes apparent. The memory effect arises due to the fact that the central node's fragility can be influenced by the previous attacks. The effect can either enhance or diminish the central node's fragility, depending on the value of the parameter $b$. In this case, we examine what happens when node 2 fails at the second time step. If the central node successfully withstands the attack from the failure of node 2, the process continues similarly. This sequential process allows us to observe the dynamics of cascading failures and how the interplay between node failures, dependence strengths, and memory effects shapes the overall robustness of the system.

\section{Results}
\subsection{The survival probability of a node}

To solve the probability $\hat{T}$ that a random node survives at the end of the cascading failure process, we first introduce an auxiliary parameter $T$, which represents the final survival probability of a random node reached by a random edge. Since a random node can survive only if none of its neighbors cause it to fail, thus the probability $\hat{T}$ can be expressed as
\begin{equation}
\label{eq:T_hat}
    \hat{T} = p \sum_{k}p(k)\sum_{n=0}^k \binom{k}{n} T ^ {k-n} (1-T) ^ n \prod_{i=0}^n[1-f(\alpha,b,i)].
\end{equation}

Further, a random node reached by a random edge survives only if none of the remaining neighboring nodes cause it to fail, thus the probability $T$ can be written as 
\begin{equation}
\label{eq:T}
    T = p \sum_{k}\frac{kp(k)}{\langle k \rangle}\sum_{n=0}^{k-1} \binom{k-1}{n} T^ {k-n-1} (1-T) ^ n \prod_{i=0}^n[1-f(\alpha,b,i)].
\end{equation}

We can define the following equation
 \begin{equation}
    h(T,p)=0
 \end{equation}
 with the function $h(T,p)$ defined as
  \begin{equation}
  \label{eq:h_T}
    h(T,p)=p \sum_{k}\frac{kp(k)}{\langle k \rangle}\sum_{n=0}^{k-1} \binom{k-1}{n} T^ {k-n-1} (1-T) ^ n \prod_{i=0}^n[1-f(\alpha,b,i)] - T.
 \end{equation}

 \begin{figure}[htb]
    \centering
    \includegraphics[width=\linewidth]{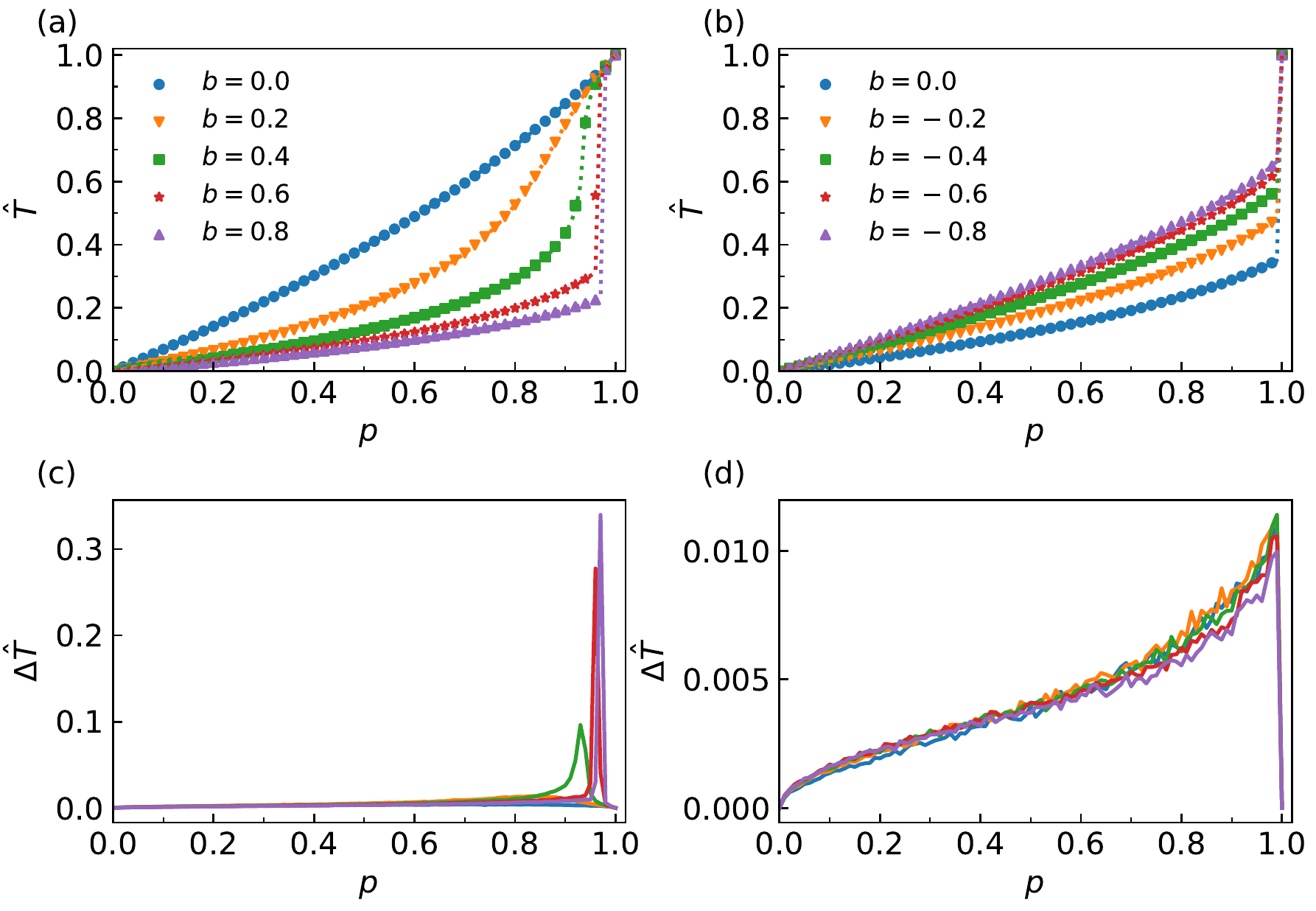}
    \caption{The final fraction $\hat{T}$ of survival nodes versus control parameter $p$ for different $\alpha$ and $b$, where the markers represent simulation results, and dashed lines represent theoretical predictions. (a) The results for $\alpha = 0.1$; (b) The results for $\alpha = 0.4$. In the two panels, the average degree is $\langle k \rangle = 4$, and a network size is $N=10^4$.}
    \label{fig:fig2}
\end{figure}

The critical point $\hat{T}_c^I$ and its corresponding auxiliary parameter $T_c^{I}$ for first-order phase transitions are determined by the control parameter $p_c^{I}$. Hence, we can first solve the following equations

 \begin{equation} \label{eq:funs}
 \left \{
\begin{aligned}
        \partial_T h \Big(p_c^I,T_{c}^{I}\Big) =0,\\
    h\Big(p_c^I,T_{c}^{I}\Big) = 0.
\end{aligned}
\right.
 \end{equation}

After obtaining the values of $(p_c^I,T_{c}^{I})$ from the previous equations, we can plug them into Eq.\ref{eq:T_hat} to calculate the critical order parameter $\hat{T}_c^I$.

Assuming that initially no nodes are removed ($p_c^I=1$) and the system remains intact ($T_c^I=1$), we can determine the critical value of the dependency strength $\alpha_c$ by analyzing the behavior of $\partial_T h (1,1)$. By observing that $\partial_T h (1,1) = 0$, we can determine the critical value $\alpha_c$ at which the system undergoes a phase transition or critical point. Since the degree follows Poisson distribution, by plugging $(p_c^I,T_c^I)=(1,1)$ into Eq.\ref{eq:funs}, we can get

\begin{equation}
    \alpha_c = \frac{1}{\langle k \rangle}.
\end{equation}




On the one hand, when $\alpha > \alpha_c$, and at least one node is removed initially, a first-order phase transition will always occur regardless of the value of $b$ (see Fig.\ref{fig:fig2}(b)). This implies that the system will undergo a sudden and significant change from $\hat{T}_{c1}$ to $\hat{T}_{c2}$ as $p$ varies from 0 to 1. 

A special case is when $b$ tends to $-\infty$, the node becomes completely resilient against subsequent attacks. In this case, Eq.\ref{eq:T} simplifies to
\begin{equation}
    T = p ( \alpha \sum_{k}\frac{kp(k)}{\langle k \rangle} T^{k-1} + 1-\alpha ).
\end{equation}
Thus, when $p=1$, the upper bound of $\hat{T}_{c1}$ will be approximately 0.74.

\begin{figure}[h]
    \centering
    \includegraphics[width=\linewidth]{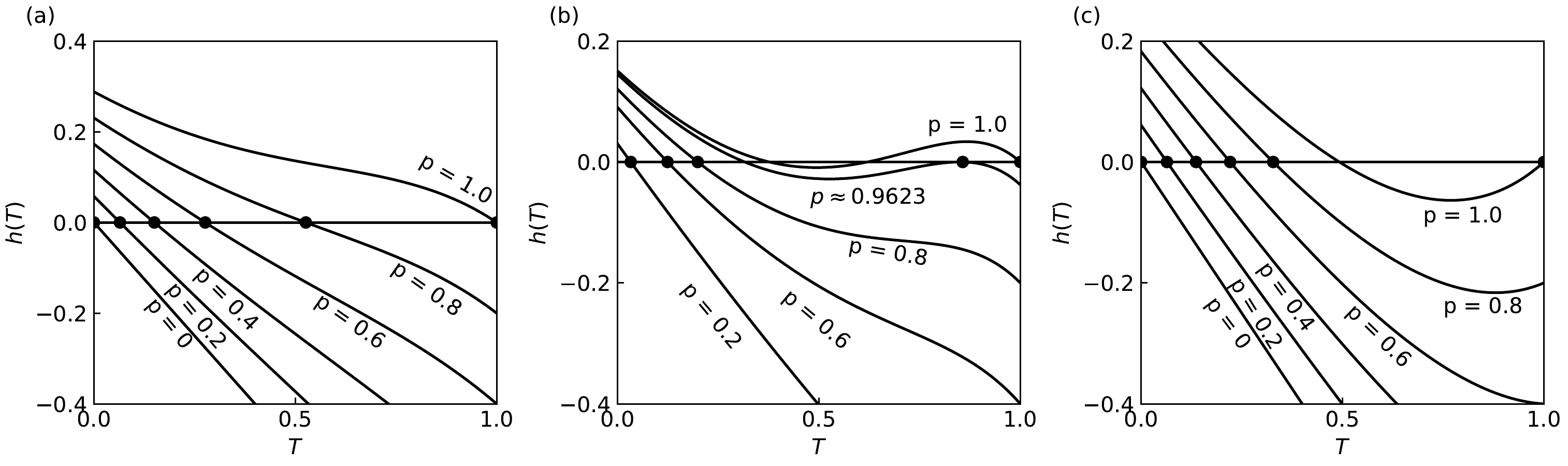}
    \caption{The function $h(T)$ for random graphs with different values of $p$, where the black dots on the horizontal axis are the solutions of $h(T)=0$. 
 (a) The results for $\langle k \rangle = 4$, $\alpha = 0.1$, and $b = 0.2$; (b) The results for $\langle k \rangle = 4$, $\alpha = 0.1$, and $b = 0.6$; (c) The results for $\langle k \rangle = 4$, $\alpha = 0.4$, and $b = -0.2$.}
    \label{fig:fig3}
\end{figure}

On the other hand, when $\alpha < \alpha_c$, a smaller value of $b$ (e.g. $b=0.2$) can compensate the attacks from the dependency strength to some extent to avoid abrupt collapse of the system, i.e. leading to a continuous increase in the final fraction $\hat{T}$ of surviving nodes as the initial removal fraction $p$ varies from 0 to 1 (see Fig.\ref{fig:fig2}(a)). This behavior can be also confirmed by the curve of $h(T)$ with $b = 0.2$ in Fig.\ref{fig:fig3}(a). However, when $b$ surpasses a critical value (e.g. $b=0.6$), the system becomes more vulnerable to sudden collapse from $T_{c2}$ to $T_{c1}$ at a critical value of $p_c \approx 0.9623$ as $p$ decreases from 1 to 0 (confirmed by Fig.\ref{fig:fig3}(b)). 

Another special case is when $b$ approaches $+\infty$, only the strength of the first wave of attack is $\alpha$, while the strengths of subsequent attacks are fixed at 1. In this case, the critical value of $p_c$ increases to approximately 0.983, and the corresponding value of $\hat{T}_{c1}$ is around 0.122.



\subsection{The size of the giant component}
To address the fraction $S$ of nodes in the final giant component, we introduce an auxiliary variable $R$, which represents the probability of a random node reached by a random edge being not connected to the giant component. Thus the probability $R$ can be expressed as 
\begin{equation}\label{eq:r}
    R = 1-T+T\sum_{k}\frac{kp(k)}{\langle k \rangle} R^{k-1}.
\end{equation}

Thus the order parameter $S$ can be given by

\begin{equation}
    S = \hat{T}\sum_{k}p(k)(1-R^k).
\end{equation}

\begin{figure}[htb]
    \centering
    \includegraphics[width=\linewidth]{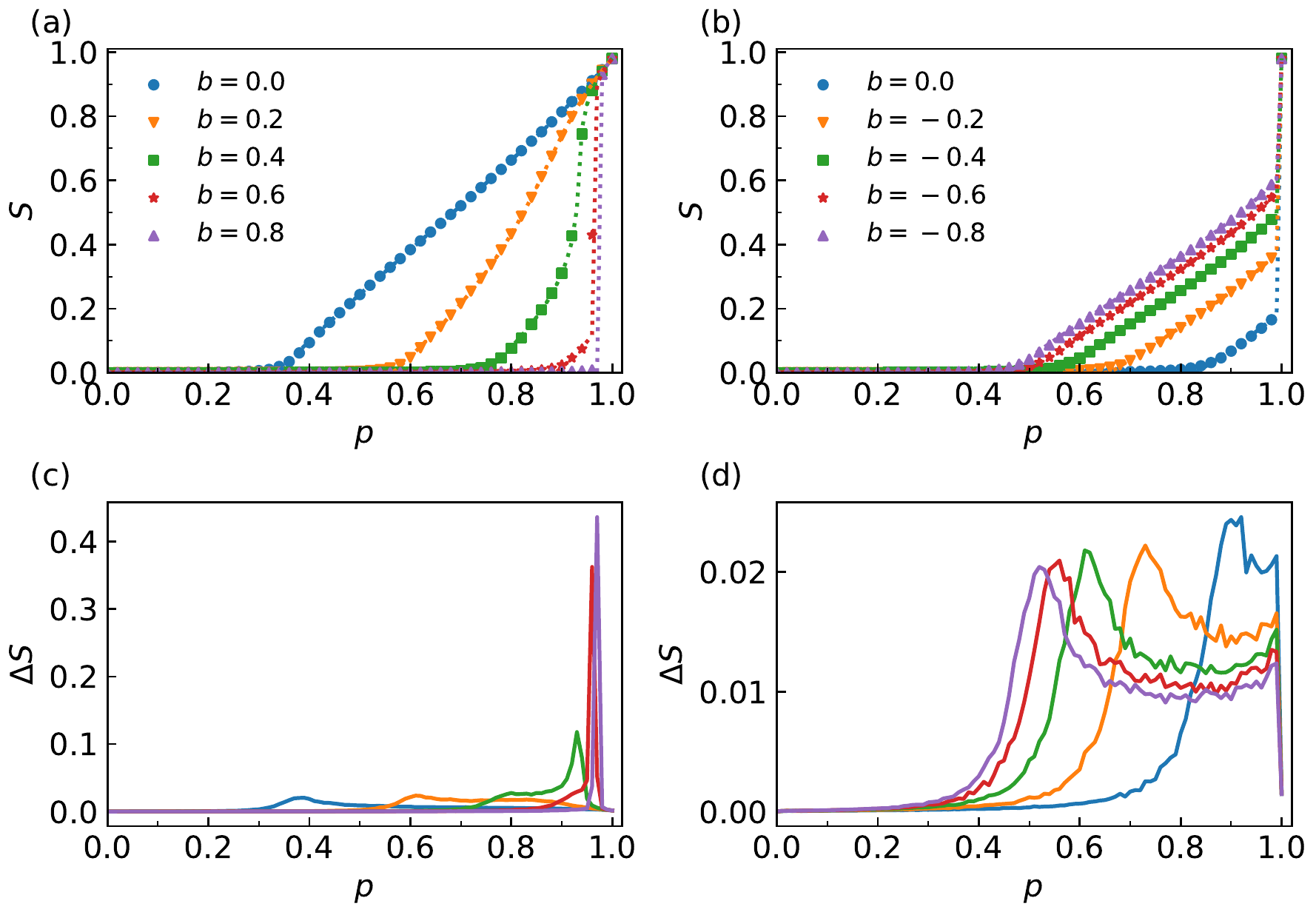}
    \caption{ The final relative size $S$ of the giant component as a function of the node reserving probability $p$ for $\alpha = 0.1$ and $\alpha = 0.4$, respectively. The network size is $N = 10^4$ and $\langle k \rangle = 4$. The markers represent simulation results, and the dashed lines represent theoretical predictions. }
    \label{fig:fig4}
\end{figure}

Since the nodes in the giant component are part of the surviving nodes, the drastic change of $\hat{T}$ will also lead to a sudden variation in $S$ at the changing point. On the one hand, for a lower value of $\alpha$ and $b$, the abrupt change of $\hat{T}$ does not exist, and the system exhibits a single second-order phase transition (see Fig.\ref{fig:fig4}(a)). 

We can define the following equation
 \begin{equation} \label{eq:g}
    g(R,T)=0
 \end{equation}
 with the function $g(R,T)$ defined as
  \begin{equation}
  \label{eq:h_T}
    g(R,T) = 1-T+T\sum_{k}\frac{kp(k)}{\langle k \rangle} R^{k-1} - R.
 \end{equation}

The critical point $S_c^{II}$ and its corresponding auxiliary parameter $R$ for second-order phase transitions are determined by the control parameter $p_c^{II}$. Since the trivial solution of Eq.\ref{eq:g} is $R=1$, the critical point $T_c^{II}$ of a second-order phase transition satisfies the following equation

\begin{equation}
    \partial_T g(1,T_c^{II}) = 0.
\end{equation}

Thus we could have 
\begin{equation}
\label{eq:T_c}
    T_c^{II}  = \frac{1}{\langle k \rangle}.
\end{equation}

By Eq.\ref{eq:T_hat}, the critical control parameter $p_c ^ {II}$ in second-order phase transition can be given  as 
\begin{equation}\label{eq:pc2}
    p_c ^ {II} = \frac{T_c^{II}}{\sum\limits_{k}\frac{kp(k)}{\langle k \rangle}\sum\limits_{n=0}^{k-1} \binom{k-1}{n} (T_c^{II})^ {k-n-1} (1-T_c^{II}) ^ n \prod\limits_{i=0}^n[1-f(\alpha,b,i)]} . 
\end{equation}

On the other hand, as $b$ increases,  a first-order phase transition with a critical value of $S_c^I$ at $p_c^{I}$ occurs (see Fig.\ref{fig:fig4}(b)). This situation can be further divided into two cases. One is that a double phase transition, i.e. a second-order and a first-order phase transition, will occur successively if the value of $T_{c1}$ is greater than the critical value of $T_{c}^{II}$ that ensures the emergence of the giant component. The other case is that as $b$ is quite large, $T_{c1}$ is less than $T_{c}^{II}$ and the second-order phase transition disappears. 



\subsection{The robustness of the system}
It has been observed that both $\alpha$ and $b$ play significant roles in determining the behavior of phase transitions and the robustness of the network.

\begin{figure}[htb]
    \centering
    \includegraphics[width=\linewidth]{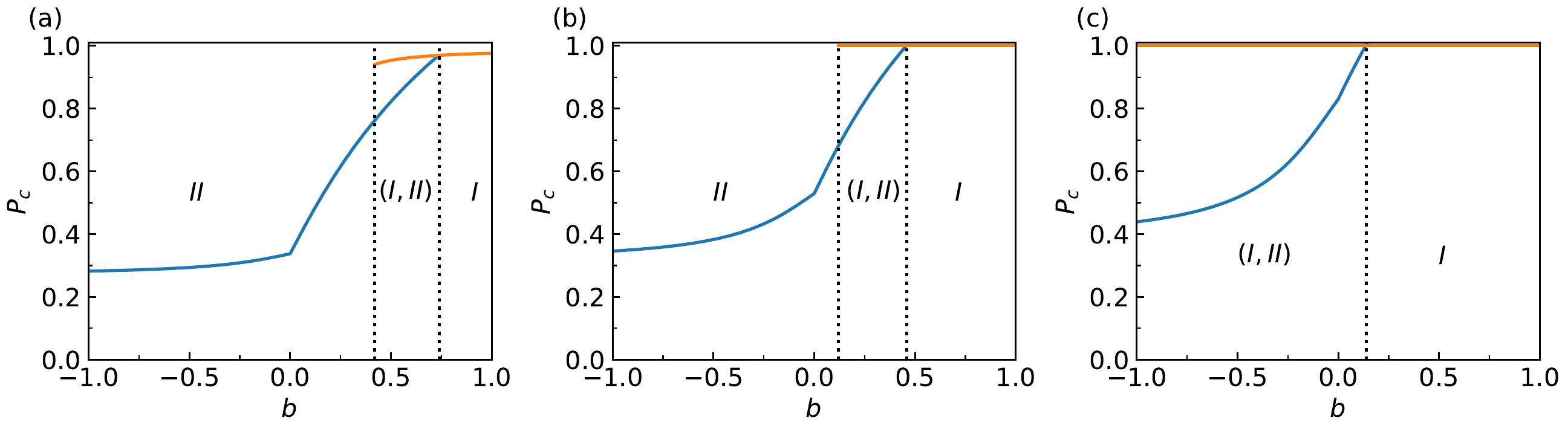}
    \caption{The theoretical results for the phase transition point $p_c$ as functions of $b$. (a-c) These show that the phase transition point $p_c ^ {I}$ or $p_c ^ {II}$ versus $b$ for $\alpha=0.1$, $\alpha=0.25$ and $\alpha=0.4$, respectively. The mean degree of the network is $\langle k \rangle = 4$ for the three panels. }
    \label{fig:fig5}
\end{figure}

When $\alpha$ is smaller than $1/{\langle k \rangle}$, as exemplified by $\alpha = 0.1$, the system exhibits a continuous emergence of the giant component in the region $II$. Increasing $b$ within the $(I, II)$ region leads to a double-phase transition, where the system first undergoes a second-order phase transition followed by a first-order phase transition. In region $I$, only a single first-order phase transition is observed, and the corresponding $p_c ^ {I}$ gradually increases with higher values of $b$ (see Fig.\ref{fig:fig5}(a)). On the other hand, when $\alpha = 1/{\langle k \rangle} = 0.25$, the system undergoes a first-order phase transition precisely at $p_c ^ {I} = 1$ in both regions $(I, II)$ and $I$ (see Fig.\ref{fig:fig5}(b)). In the case of $\alpha$ being greater than $1/{\langle k \rangle}$, for instance, $\alpha = 0.4$, two distinct regions are observed: the double-phase transition region $(I, II)$ and the first-order phase transition region $I$. Irrespective of the value of $b$, a first-order phase transition occurs at $p_c ^ {I} = 1$ (see Fig.\ref{fig:fig5}(c)).

\begin{figure}[htb]
    \centering
    \includegraphics[width=\linewidth]{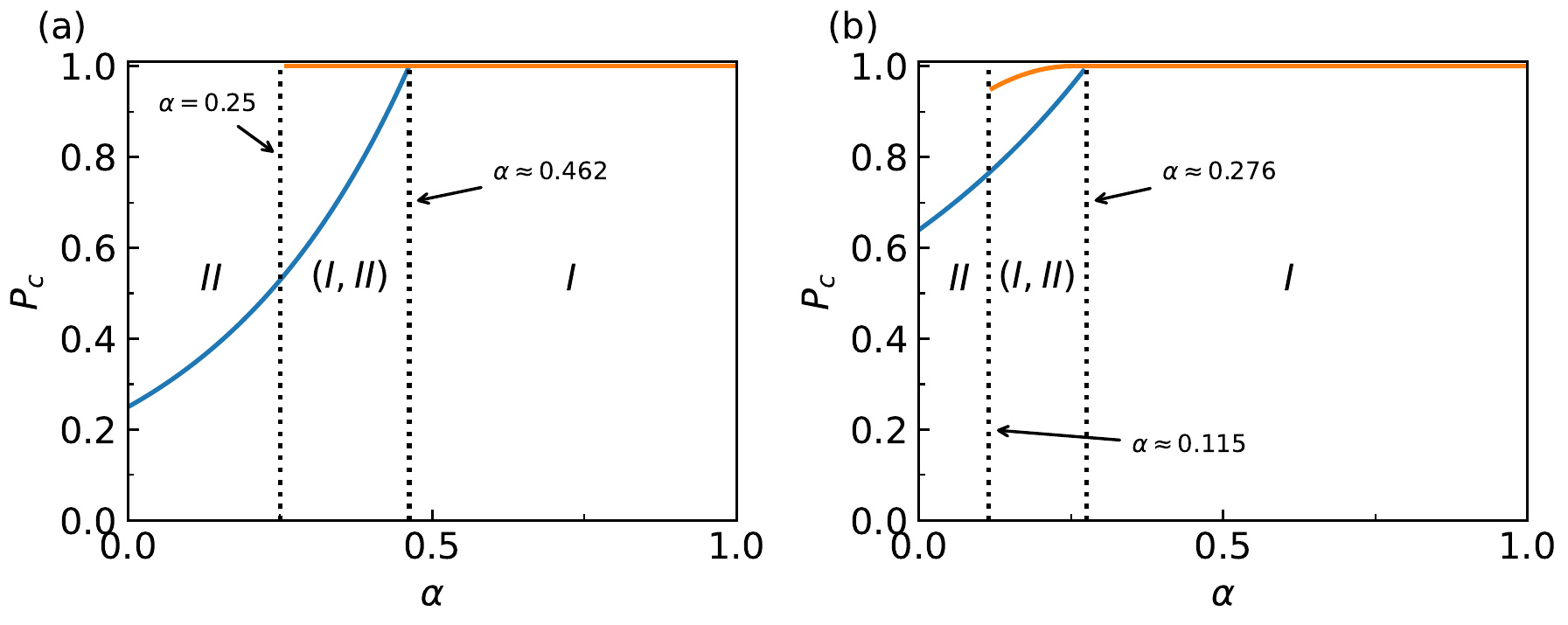}
    \caption{The theoretical results for the phase transition point $p_c$ as functions of $\alpha$. (a-b) These show that the phase transition point $p_c ^ {I}$ or $p_c ^ {II}$ versus $\alpha$ for $b=0$ and $b=0.4$. The mean degree of the network is $\langle k \rangle = 4$ for the two panels. }
    \label{fig:fig6}
\end{figure}

Furthermore, when $b=0$, the behavior of the system is influenced by $\alpha$. For $\alpha$ values smaller than $1/{\langle k \rangle} = 0.25$, the giant component emerges continuously as the percolation threshold is crossed. However, as $\alpha$ increases, reaching approximately 0.462, the system undergoes a double-phase transition in the region $(I, II)$. Beyond this threshold, the second-order phase transition disappears, and the system transfers to a single first-order phase transition in the region $I$ (see Fig.\ref{fig:fig6}(a)). When $b>0$, such as $b=0.4$, the critical point $\alpha_c$ decreases to approximately 0.115. Further increasing $\alpha$ to around 0.276 within region $(I, II)$ leads to a double-phase transition, which then transfers to a single first-order phase transition in the region $I$ (see Fig.\ref{fig:fig6}(b)).


\newpage
\section{Discussion and Conclusion}
The purpose of this paper is to investigate cascading failures in complex networks, focusing on the dependencies between nodes and memory effects. A model is proposed to simulate cascading failures with non-Markovian node dependencies, where a node's failure probability depends on both its current state and past attack history. The study combines theoretical analysis and numerical simulations to explore the influence of dependency strength, memory effects, and network structure on cascading failures. Different types of phase transitions and conditions for system robustness are identified. 

The paper uncovers subtle relationships between dependency strength, memory effects, and network structure, and their impact on system phase transitions and stability. The findings reveal that when the dependency strength exceeds the inverse of the average degree, the system undergoes sudden collapse even with the removal of an infinitesimally small fraction of nodes, irrespective of the memory effect. When the dependency strength is below the inverse of the average degree, a smaller memory effect can partially offset the negative impact of dependency strength, resulting in a gradual increase in the fraction of surviving nodes. However, exceeding a critical threshold, a larger memory effect renders the system more vulnerable, leading to abrupt collapse. The paper also investigates the influence of network structure on phase transition types and critical points, providing theoretical analyses. 

Moreover, the study suggests the potential use of memory effects to strengthen or weaken dependencies between nodes, thus enhancing network robustness. The authors propose memory effects as an adaptive regulation mechanism, enabling nodes to adjust their response to neighboring node failures following attacks. Positive memory effects increase dependencies, making nodes more susceptible to neighboring node failures, while negative memory effects reduce dependencies, enhancing node resilience. By appropriately designing and adjusting memory effects, networks can exhibit improved robustness and adaptability in the face of attacks or failures.

In conclusion, this paper has significant implications for assessing and managing risks and vulnerabilities in complex systems. It also raises intriguing avenues for future research, such as considering different types or directions of dependencies, networks with diverse distributions or topological structures, and dynamic or adaptive memory effects. These investigations can further enrich the theoretical framework and broaden the application domains of cascading failures, providing valuable insights for ensuring the stability and security of complex networks.

\begin{acknowledgements}
    This work is supported by the National Natural Science Foundation of China (61773148) and the Entrepreneurship and Innovation Project of High Level Returned Overseas Scholar in Hangzhou.
\end{acknowledgements}

\section{References}
\bibliography{aipsamp}

\begin{thebibliography}{31}%
\makeatletter
\providecommand \@ifxundefined [1]{%
 \@ifx{#1\undefined}
}%
\providecommand \@ifnum [1]{%
 \ifnum #1\expandafter \@firstoftwo
 \else \expandafter \@secondoftwo
 \fi
}%
\providecommand \@ifx [1]{%
 \ifx #1\expandafter \@firstoftwo
 \else \expandafter \@secondoftwo
 \fi
}%
\providecommand \natexlab [1]{#1}%
\providecommand \enquote  [1]{``#1''}%
\providecommand \bibnamefont  [1]{#1}%
\providecommand \bibfnamefont [1]{#1}%
\providecommand \citenamefont [1]{#1}%
\providecommand \href@noop [0]{\@secondoftwo}%
\providecommand \href [0]{\begingroup \@sanitize@url \@href}%
\providecommand \@href[1]{\@@startlink{#1}\@@href}%
\providecommand \@@href[1]{\endgroup#1\@@endlink}%
\providecommand \@sanitize@url [0]{\catcode `\\12\catcode `\$12\catcode
  `\&12\catcode `\#12\catcode `\^12\catcode `\_12\catcode `\%12\relax}%
\providecommand \@@startlink[1]{}%
\providecommand \@@endlink[0]{}%
\providecommand \url  [0]{\begingroup\@sanitize@url \@url }%
\providecommand \@url [1]{\endgroup\@href {#1}{\urlprefix }}%
\providecommand \urlprefix  [0]{URL }%
\providecommand \Eprint [0]{\href }%
\providecommand \doibase [0]{http://dx.doi.org/}%
\providecommand \selectlanguage [0]{\@gobble}%
\providecommand \bibinfo  [0]{\@secondoftwo}%
\providecommand \bibfield  [0]{\@secondoftwo}%
\providecommand \translation [1]{[#1]}%
\providecommand \BibitemOpen [0]{}%
\providecommand \bibitemStop [0]{}%
\providecommand \bibitemNoStop [0]{.\EOS\space}%
\providecommand \EOS [0]{\spacefactor3000\relax}%
\providecommand \BibitemShut  [1]{\csname bibitem#1\endcsname}%
\let\auto@bib@innerbib\@empty
\bibitem [{\citenamefont {Valdez}\ \emph {et~al.}(2020)\citenamefont {Valdez},
  \citenamefont {Shekhtman}, \citenamefont {La~Rocca}, \citenamefont {Zhang},
  \citenamefont {Buldyrev}, \citenamefont {Trunfio}, \citenamefont
  {Braunstein},\ and\ \citenamefont {Havlin}}]{valdez2020cascading}%
  \BibitemOpen
  \bibfield  {author} {\bibinfo {author} {\bibfnamefont {L.~D.}\ \bibnamefont
  {Valdez}}, \bibinfo {author} {\bibfnamefont {L.}~\bibnamefont {Shekhtman}},
  \bibinfo {author} {\bibfnamefont {C.~E.}\ \bibnamefont {La~Rocca}}, \bibinfo
  {author} {\bibfnamefont {X.}~\bibnamefont {Zhang}}, \bibinfo {author}
  {\bibfnamefont {S.~V.}\ \bibnamefont {Buldyrev}}, \bibinfo {author}
  {\bibfnamefont {P.~A.}\ \bibnamefont {Trunfio}}, \bibinfo {author}
  {\bibfnamefont {L.~A.}\ \bibnamefont {Braunstein}}, \ and\ \bibinfo {author}
  {\bibfnamefont {S.}~\bibnamefont {Havlin}},\ }\bibfield  {title} {\enquote
  {\bibinfo {title} {Cascading failures in complex networks},}\ }\href@noop {}
  {\bibfield  {journal} {\bibinfo  {journal} {Journal of Complex Networks}\
  }\textbf {\bibinfo {volume} {8}},\ \bibinfo {pages} {cnaa013} (\bibinfo
  {year} {2020})}\BibitemShut {NoStop}%
\bibitem [{\citenamefont {Duenas-Osorio}\ and\ \citenamefont
  {Vemuru}(2009)}]{duenas2009cascading}%
  \BibitemOpen
  \bibfield  {author} {\bibinfo {author} {\bibfnamefont {L.}~\bibnamefont
  {Duenas-Osorio}}\ and\ \bibinfo {author} {\bibfnamefont {S.~M.}\ \bibnamefont
  {Vemuru}},\ }\bibfield  {title} {\enquote {\bibinfo {title} {Cascading
  failures in complex infrastructure systems},}\ }\href@noop {} {\bibfield
  {journal} {\bibinfo  {journal} {Structural Safety}\ }\textbf {\bibinfo
  {volume} {31}},\ \bibinfo {pages} {157--167} (\bibinfo {year}
  {2009})}\BibitemShut {NoStop}%
\bibitem [{\citenamefont {Crucitti}, \citenamefont {Latora},\ and\
  \citenamefont {Marchiori}(2004)}]{crucitti2004model}%
  \BibitemOpen
  \bibfield  {author} {\bibinfo {author} {\bibfnamefont {P.}~\bibnamefont
  {Crucitti}}, \bibinfo {author} {\bibfnamefont {V.}~\bibnamefont {Latora}}, \
  and\ \bibinfo {author} {\bibfnamefont {M.}~\bibnamefont {Marchiori}},\
  }\bibfield  {title} {\enquote {\bibinfo {title} {Model for cascading failures
  in complex networks},}\ }\href@noop {} {\bibfield  {journal} {\bibinfo
  {journal} {Physical Review E}\ }\textbf {\bibinfo {volume} {69}},\ \bibinfo
  {pages} {045104} (\bibinfo {year} {2004})}\BibitemShut {NoStop}%
\bibitem [{\citenamefont {Sch{\"a}fer}\ \emph {et~al.}(2018)\citenamefont
  {Sch{\"a}fer}, \citenamefont {Witthaut}, \citenamefont {Timme},\ and\
  \citenamefont {Latora}}]{schafer2018dynamically}%
  \BibitemOpen
  \bibfield  {author} {\bibinfo {author} {\bibfnamefont {B.}~\bibnamefont
  {Sch{\"a}fer}}, \bibinfo {author} {\bibfnamefont {D.}~\bibnamefont
  {Witthaut}}, \bibinfo {author} {\bibfnamefont {M.}~\bibnamefont {Timme}}, \
  and\ \bibinfo {author} {\bibfnamefont {V.}~\bibnamefont {Latora}},\
  }\bibfield  {title} {\enquote {\bibinfo {title} {Dynamically induced
  cascading failures in power grids},}\ }\href@noop {} {\bibfield  {journal}
  {\bibinfo  {journal} {Nature Communications}\ }\textbf {\bibinfo {volume}
  {9}},\ \bibinfo {pages} {1975} (\bibinfo {year} {2018})}\BibitemShut
  {NoStop}%
\bibitem [{\citenamefont {Pahwa}, \citenamefont {Scoglio},\ and\ \citenamefont
  {Scala}(2014)}]{pahwa2014abruptness}%
  \BibitemOpen
  \bibfield  {author} {\bibinfo {author} {\bibfnamefont {S.}~\bibnamefont
  {Pahwa}}, \bibinfo {author} {\bibfnamefont {C.}~\bibnamefont {Scoglio}}, \
  and\ \bibinfo {author} {\bibfnamefont {A.}~\bibnamefont {Scala}},\ }\bibfield
   {title} {\enquote {\bibinfo {title} {Abruptness of cascade failures in power
  grids},}\ }\href@noop {} {\bibfield  {journal} {\bibinfo  {journal}
  {Scientific Reports}\ }\textbf {\bibinfo {volume} {4}},\ \bibinfo {pages}
  {1--9} (\bibinfo {year} {2014})}\BibitemShut {NoStop}%
\bibitem [{\citenamefont {Candelieri}\ \emph {et~al.}(2019)\citenamefont
  {Candelieri}, \citenamefont {Galuzzi}, \citenamefont {Giordani},\ and\
  \citenamefont {Archetti}}]{candelieri2019vulnerability}%
  \BibitemOpen
  \bibfield  {author} {\bibinfo {author} {\bibfnamefont {A.}~\bibnamefont
  {Candelieri}}, \bibinfo {author} {\bibfnamefont {B.~G.}\ \bibnamefont
  {Galuzzi}}, \bibinfo {author} {\bibfnamefont {I.}~\bibnamefont {Giordani}}, \
  and\ \bibinfo {author} {\bibfnamefont {F.}~\bibnamefont {Archetti}},\
  }\bibfield  {title} {\enquote {\bibinfo {title} {Vulnerability of public
  transportation networks against directed attacks and cascading failures},}\
  }\href@noop {} {\bibfield  {journal} {\bibinfo  {journal} {Public Transport}\
  }\textbf {\bibinfo {volume} {11}},\ \bibinfo {pages} {27--49} (\bibinfo
  {year} {2019})}\BibitemShut {NoStop}%
\bibitem [{\citenamefont {Li}\ \emph {et~al.}(2022)\citenamefont {Li},
  \citenamefont {Lv}, \citenamefont {Lang},\ and\ \citenamefont
  {Chen}}]{li2022exploring}%
  \BibitemOpen
  \bibfield  {author} {\bibinfo {author} {\bibfnamefont {X.}~\bibnamefont
  {Li}}, \bibinfo {author} {\bibfnamefont {B.}~\bibnamefont {Lv}}, \bibinfo
  {author} {\bibfnamefont {B.}~\bibnamefont {Lang}}, \ and\ \bibinfo {author}
  {\bibfnamefont {Q.}~\bibnamefont {Chen}},\ }\bibfield  {title} {\enquote
  {\bibinfo {title} {Exploring the cascading failure in taxi transportation
  networks},}\ }\href@noop {} {\bibfield  {journal} {\bibinfo  {journal}
  {Sustainability}\ }\textbf {\bibinfo {volume} {14}},\ \bibinfo {pages}
  {13221} (\bibinfo {year} {2022})}\BibitemShut {NoStop}%
\bibitem [{\citenamefont {Wu}\ \emph {et~al.}(2022)\citenamefont {Wu},
  \citenamefont {Yang}, \citenamefont {Jiang}, \citenamefont {Zhan},
  \citenamefont {Liao}, \citenamefont {Xu}, \citenamefont {Fan},\ and\
  \citenamefont {Liang}}]{wu2022cascading}%
  \BibitemOpen
  \bibfield  {author} {\bibinfo {author} {\bibfnamefont {F.}~\bibnamefont
  {Wu}}, \bibinfo {author} {\bibfnamefont {J.}~\bibnamefont {Yang}}, \bibinfo
  {author} {\bibfnamefont {H.}~\bibnamefont {Jiang}}, \bibinfo {author}
  {\bibfnamefont {X.}~\bibnamefont {Zhan}}, \bibinfo {author} {\bibfnamefont
  {S.}~\bibnamefont {Liao}}, \bibinfo {author} {\bibfnamefont {J.}~\bibnamefont
  {Xu}}, \bibinfo {author} {\bibfnamefont {H.}~\bibnamefont {Fan}}, \ and\
  \bibinfo {author} {\bibfnamefont {J.}~\bibnamefont {Liang}},\ }\bibfield
  {title} {\enquote {\bibinfo {title} {Cascading failure in coupled networks of
  transportation and power grid},}\ }\href@noop {} {\bibfield  {journal}
  {\bibinfo  {journal} {International Journal of Electrical Power \& Energy
  Systems}\ }\textbf {\bibinfo {volume} {140}},\ \bibinfo {pages} {108058}
  (\bibinfo {year} {2022})}\BibitemShut {NoStop}%
\bibitem [{\citenamefont {Ramirez}, \citenamefont {van~den Hoven},\ and\
  \citenamefont {Bauso}(2023)}]{ramirez2023stochastic}%
  \BibitemOpen
  \bibfield  {author} {\bibinfo {author} {\bibfnamefont {S.}~\bibnamefont
  {Ramirez}}, \bibinfo {author} {\bibfnamefont {M.}~\bibnamefont {van~den
  Hoven}}, \ and\ \bibinfo {author} {\bibfnamefont {D.}~\bibnamefont {Bauso}},\
  }\bibfield  {title} {\enquote {\bibinfo {title} {A stochastic model for
  cascading failures in financial networks},}\ }\href@noop {} {\bibfield
  {journal} {\bibinfo  {journal} {IEEE Transactions on Control of Network
  Systems}\ } (\bibinfo {year} {2023})}\BibitemShut {NoStop}%
\bibitem [{\citenamefont {Huang}\ \emph {et~al.}(2013)\citenamefont {Huang},
  \citenamefont {Vodenska}, \citenamefont {Havlin},\ and\ \citenamefont
  {Stanley}}]{huang2013cascading}%
  \BibitemOpen
  \bibfield  {author} {\bibinfo {author} {\bibfnamefont {X.}~\bibnamefont
  {Huang}}, \bibinfo {author} {\bibfnamefont {I.}~\bibnamefont {Vodenska}},
  \bibinfo {author} {\bibfnamefont {S.}~\bibnamefont {Havlin}}, \ and\ \bibinfo
  {author} {\bibfnamefont {H.~E.}\ \bibnamefont {Stanley}},\ }\bibfield
  {title} {\enquote {\bibinfo {title} {Cascading failures in bi-partite graphs:
  model for systemic risk propagation},}\ }\href@noop {} {\bibfield  {journal}
  {\bibinfo  {journal} {Scientific Reports}\ }\textbf {\bibinfo {volume} {3}},\
  \bibinfo {pages} {1219} (\bibinfo {year} {2013})}\BibitemShut {NoStop}%
\bibitem [{\citenamefont {Yi}\ \emph {et~al.}(2015)\citenamefont {Yi},
  \citenamefont {Bao}, \citenamefont {Jiang},\ and\ \citenamefont
  {Xue}}]{yi2015modeling}%
  \BibitemOpen
  \bibfield  {author} {\bibinfo {author} {\bibfnamefont {C.}~\bibnamefont
  {Yi}}, \bibinfo {author} {\bibfnamefont {Y.}~\bibnamefont {Bao}}, \bibinfo
  {author} {\bibfnamefont {J.}~\bibnamefont {Jiang}}, \ and\ \bibinfo {author}
  {\bibfnamefont {Y.}~\bibnamefont {Xue}},\ }\bibfield  {title} {\enquote
  {\bibinfo {title} {Modeling cascading failures with the crisis of trust in
  social networks},}\ }\href@noop {} {\bibfield  {journal} {\bibinfo  {journal}
  {Physica A: Statistical Mechanics and its Applications}\ }\textbf {\bibinfo
  {volume} {436}},\ \bibinfo {pages} {256--271} (\bibinfo {year}
  {2015})}\BibitemShut {NoStop}%
\bibitem [{\citenamefont {Motter}\ and\ \citenamefont
  {Lai}(2002)}]{motter2002cascade}%
  \BibitemOpen
  \bibfield  {author} {\bibinfo {author} {\bibfnamefont {A.~E.}\ \bibnamefont
  {Motter}}\ and\ \bibinfo {author} {\bibfnamefont {Y.-C.}\ \bibnamefont
  {Lai}},\ }\bibfield  {title} {\enquote {\bibinfo {title} {Cascade-based
  attacks on complex networks},}\ }\href@noop {} {\bibfield  {journal}
  {\bibinfo  {journal} {Physical Review E}\ }\textbf {\bibinfo {volume} {66}},\
  \bibinfo {pages} {065102} (\bibinfo {year} {2002})}\BibitemShut {NoStop}%
\bibitem [{\citenamefont {Baxter}\ \emph {et~al.}(2010)\citenamefont {Baxter},
  \citenamefont {Dorogovtsev}, \citenamefont {Goltsev},\ and\ \citenamefont
  {Mendes}}]{baxter2010bootstrap}%
  \BibitemOpen
  \bibfield  {author} {\bibinfo {author} {\bibfnamefont {G.~J.}\ \bibnamefont
  {Baxter}}, \bibinfo {author} {\bibfnamefont {S.~N.}\ \bibnamefont
  {Dorogovtsev}}, \bibinfo {author} {\bibfnamefont {A.~V.}\ \bibnamefont
  {Goltsev}}, \ and\ \bibinfo {author} {\bibfnamefont {J.~F.}\ \bibnamefont
  {Mendes}},\ }\bibfield  {title} {\enquote {\bibinfo {title} {Bootstrap
  percolation on complex networks},}\ }\href@noop {} {\bibfield  {journal}
  {\bibinfo  {journal} {Physical Review E}\ }\textbf {\bibinfo {volume} {82}},\
  \bibinfo {pages} {011103} (\bibinfo {year} {2010})}\BibitemShut {NoStop}%
\bibitem [{\citenamefont {Holme}\ and\ \citenamefont
  {Kim}(2002)}]{holme2002vertex}%
  \BibitemOpen
  \bibfield  {author} {\bibinfo {author} {\bibfnamefont {P.}~\bibnamefont
  {Holme}}\ and\ \bibinfo {author} {\bibfnamefont {B.~J.}\ \bibnamefont
  {Kim}},\ }\bibfield  {title} {\enquote {\bibinfo {title} {Vertex overload
  breakdown in evolving networks},}\ }\href@noop {} {\bibfield  {journal}
  {\bibinfo  {journal} {Physical Review E}\ }\textbf {\bibinfo {volume} {65}},\
  \bibinfo {pages} {066109} (\bibinfo {year} {2002})}\BibitemShut {NoStop}%
\bibitem [{\citenamefont {Holme}(2002)}]{holme2002edge}%
  \BibitemOpen
  \bibfield  {author} {\bibinfo {author} {\bibfnamefont {P.}~\bibnamefont
  {Holme}},\ }\bibfield  {title} {\enquote {\bibinfo {title} {Edge overload
  breakdown in evolving networks},}\ }\href@noop {} {\bibfield  {journal}
  {\bibinfo  {journal} {Physical Review E}\ }\textbf {\bibinfo {volume} {66}},\
  \bibinfo {pages} {036119} (\bibinfo {year} {2002})}\BibitemShut {NoStop}%
\bibitem [{\citenamefont {Parshani}, \citenamefont {Buldyrev},\ and\
  \citenamefont {Havlin}(2011)}]{parshani2011critical}%
  \BibitemOpen
  \bibfield  {author} {\bibinfo {author} {\bibfnamefont {R.}~\bibnamefont
  {Parshani}}, \bibinfo {author} {\bibfnamefont {S.~V.}\ \bibnamefont
  {Buldyrev}}, \ and\ \bibinfo {author} {\bibfnamefont {S.}~\bibnamefont
  {Havlin}},\ }\bibfield  {title} {\enquote {\bibinfo {title} {Critical effect
  of dependency groups on the function of networks},}\ }\href@noop {}
  {\bibfield  {journal} {\bibinfo  {journal} {Proceedings of the National
  Academy of Sciences}\ }\textbf {\bibinfo {volume} {108}},\ \bibinfo {pages}
  {1007--1010} (\bibinfo {year} {2011})}\BibitemShut {NoStop}%
\bibitem [{\citenamefont {Buldyrev}\ \emph {et~al.}(2010)\citenamefont
  {Buldyrev}, \citenamefont {Parshani}, \citenamefont {Paul}, \citenamefont
  {Stanley},\ and\ \citenamefont {Havlin}}]{buldyrev2010catastrophic}%
  \BibitemOpen
  \bibfield  {author} {\bibinfo {author} {\bibfnamefont {S.~V.}\ \bibnamefont
  {Buldyrev}}, \bibinfo {author} {\bibfnamefont {R.}~\bibnamefont {Parshani}},
  \bibinfo {author} {\bibfnamefont {G.}~\bibnamefont {Paul}}, \bibinfo {author}
  {\bibfnamefont {H.~E.}\ \bibnamefont {Stanley}}, \ and\ \bibinfo {author}
  {\bibfnamefont {S.}~\bibnamefont {Havlin}},\ }\bibfield  {title} {\enquote
  {\bibinfo {title} {Catastrophic cascade of failures in interdependent
  networks},}\ }\href@noop {} {\bibfield  {journal} {\bibinfo  {journal}
  {Nature}\ }\textbf {\bibinfo {volume} {464}},\ \bibinfo {pages} {1025--1028}
  (\bibinfo {year} {2010})}\BibitemShut {NoStop}%
\bibitem [{\citenamefont {Liu}\ \emph {et~al.}(2016)\citenamefont {Liu},
  \citenamefont {Li}, \citenamefont {Jia},\ and\ \citenamefont
  {Wang}}]{liu2016cascading}%
  \BibitemOpen
  \bibfield  {author} {\bibinfo {author} {\bibfnamefont {R.-R.}\ \bibnamefont
  {Liu}}, \bibinfo {author} {\bibfnamefont {M.}~\bibnamefont {Li}}, \bibinfo
  {author} {\bibfnamefont {C.-X.}\ \bibnamefont {Jia}}, \ and\ \bibinfo
  {author} {\bibfnamefont {B.-H.}\ \bibnamefont {Wang}},\ }\bibfield  {title}
  {\enquote {\bibinfo {title} {Cascading failures in coupled networks with both
  inner-dependency and inter-dependency links},}\ }\href@noop {} {\bibfield
  {journal} {\bibinfo  {journal} {Scientific Reports}\ }\textbf {\bibinfo
  {volume} {6}},\ \bibinfo {pages} {1--10} (\bibinfo {year}
  {2016})}\BibitemShut {NoStop}%
\bibitem [{\citenamefont {Liu}\ \emph {et~al.}(2018)\citenamefont {Liu},
  \citenamefont {Eisenberg}, \citenamefont {Seager},\ and\ \citenamefont
  {Lai}}]{liu2018weak}%
  \BibitemOpen
  \bibfield  {author} {\bibinfo {author} {\bibfnamefont {R.-R.}\ \bibnamefont
  {Liu}}, \bibinfo {author} {\bibfnamefont {D.~A.}\ \bibnamefont {Eisenberg}},
  \bibinfo {author} {\bibfnamefont {T.~P.}\ \bibnamefont {Seager}}, \ and\
  \bibinfo {author} {\bibfnamefont {Y.-C.}\ \bibnamefont {Lai}},\ }\bibfield
  {title} {\enquote {\bibinfo {title} {The “weak” interdependence of
  infrastructure systems produces mixed percolation transitions in multilayer
  networks},}\ }\href@noop {} {\bibfield  {journal} {\bibinfo  {journal}
  {Scientific reports}\ }\textbf {\bibinfo {volume} {8}},\ \bibinfo {pages}
  {2111} (\bibinfo {year} {2018})}\BibitemShut {NoStop}%
\bibitem [{\citenamefont {Watts}(2002)}]{watts2002simple}%
  \BibitemOpen
  \bibfield  {author} {\bibinfo {author} {\bibfnamefont {D.~J.}\ \bibnamefont
  {Watts}},\ }\bibfield  {title} {\enquote {\bibinfo {title} {A simple model of
  global cascades on random networks},}\ }\href@noop {} {\bibfield  {journal}
  {\bibinfo  {journal} {Proceedings of the National Academy of Sciences}\
  }\textbf {\bibinfo {volume} {99}},\ \bibinfo {pages} {5766--5771} (\bibinfo
  {year} {2002})}\BibitemShut {NoStop}%
\bibitem [{\citenamefont {Cellai}\ \emph {et~al.}(2011)\citenamefont {Cellai},
  \citenamefont {Lawlor}, \citenamefont {Dawson},\ and\ \citenamefont
  {Gleeson}}]{cellai2011tricritical}%
  \BibitemOpen
  \bibfield  {author} {\bibinfo {author} {\bibfnamefont {D.}~\bibnamefont
  {Cellai}}, \bibinfo {author} {\bibfnamefont {A.}~\bibnamefont {Lawlor}},
  \bibinfo {author} {\bibfnamefont {K.~A.}\ \bibnamefont {Dawson}}, \ and\
  \bibinfo {author} {\bibfnamefont {J.~P.}\ \bibnamefont {Gleeson}},\
  }\bibfield  {title} {\enquote {\bibinfo {title} {Tricritical point in
  heterogeneous k-core percolation},}\ }\href@noop {} {\bibfield  {journal}
  {\bibinfo  {journal} {Physical Review Letters}\ }\textbf {\bibinfo {volume}
  {107}},\ \bibinfo {pages} {175703} (\bibinfo {year} {2011})}\BibitemShut
  {NoStop}%
\bibitem [{\citenamefont {Dorogovtsev}, \citenamefont {Goltsev},\ and\
  \citenamefont {Mendes}(2006)}]{dorogovtsev2006k}%
  \BibitemOpen
  \bibfield  {author} {\bibinfo {author} {\bibfnamefont {S.~N.}\ \bibnamefont
  {Dorogovtsev}}, \bibinfo {author} {\bibfnamefont {A.~V.}\ \bibnamefont
  {Goltsev}}, \ and\ \bibinfo {author} {\bibfnamefont {J.~F.~F.}\ \bibnamefont
  {Mendes}},\ }\bibfield  {title} {\enquote {\bibinfo {title} {K-core
  organization of complex networks},}\ }\href@noop {} {\bibfield  {journal}
  {\bibinfo  {journal} {Physical Review Letters}\ }\textbf {\bibinfo {volume}
  {96}},\ \bibinfo {pages} {040601} (\bibinfo {year} {2006})}\BibitemShut
  {NoStop}%
\bibitem [{\citenamefont {Feng}\ \emph {et~al.}(2019)\citenamefont {Feng},
  \citenamefont {Cai}, \citenamefont {Tang},\ and\ \citenamefont
  {Lai}}]{feng2019equivalence}%
  \BibitemOpen
  \bibfield  {author} {\bibinfo {author} {\bibfnamefont {M.}~\bibnamefont
  {Feng}}, \bibinfo {author} {\bibfnamefont {S.-M.}\ \bibnamefont {Cai}},
  \bibinfo {author} {\bibfnamefont {M.}~\bibnamefont {Tang}}, \ and\ \bibinfo
  {author} {\bibfnamefont {Y.-C.}\ \bibnamefont {Lai}},\ }\bibfield  {title}
  {\enquote {\bibinfo {title} {Equivalence and its invalidation between
  non-markovian and markovian spreading dynamics on complex networks},}\
  }\href@noop {} {\bibfield  {journal} {\bibinfo  {journal} {Nature
  Communications}\ }\textbf {\bibinfo {volume} {10}},\ \bibinfo {pages} {3748}
  (\bibinfo {year} {2019})}\BibitemShut {NoStop}%
\bibitem [{\citenamefont {Starnini}, \citenamefont {Gleeson},\ and\
  \citenamefont {Bogu{\~n}{\'a}}(2017)}]{starnini2017equivalence}%
  \BibitemOpen
  \bibfield  {author} {\bibinfo {author} {\bibfnamefont {M.}~\bibnamefont
  {Starnini}}, \bibinfo {author} {\bibfnamefont {J.~P.}\ \bibnamefont
  {Gleeson}}, \ and\ \bibinfo {author} {\bibfnamefont {M.}~\bibnamefont
  {Bogu{\~n}{\'a}}},\ }\bibfield  {title} {\enquote {\bibinfo {title}
  {Equivalence between non-markovian and markovian dynamics in epidemic
  spreading processes},}\ }\href@noop {} {\bibfield  {journal} {\bibinfo
  {journal} {Physical Review Letters}\ }\textbf {\bibinfo {volume} {118}},\
  \bibinfo {pages} {128301} (\bibinfo {year} {2017})}\BibitemShut {NoStop}%
\bibitem [{\citenamefont {Tomovski}, \citenamefont {Basnarkov},\ and\
  \citenamefont {Abazi}(2021)}]{tomovski2021discrete}%
  \BibitemOpen
  \bibfield  {author} {\bibinfo {author} {\bibfnamefont {I.}~\bibnamefont
  {Tomovski}}, \bibinfo {author} {\bibfnamefont {L.}~\bibnamefont {Basnarkov}},
  \ and\ \bibinfo {author} {\bibfnamefont {A.}~\bibnamefont {Abazi}},\
  }\bibfield  {title} {\enquote {\bibinfo {title} {Discrete-time non-markovian
  seis model on complex networks},}\ }\href@noop {} {\bibfield  {journal}
  {\bibinfo  {journal} {IEEE Transactions on Network Science and Engineering}\
  }\textbf {\bibinfo {volume} {9}},\ \bibinfo {pages} {552--563} (\bibinfo
  {year} {2021})}\BibitemShut {NoStop}%
\bibitem [{\citenamefont {Lenzi}, \citenamefont {Yednak},\ and\ \citenamefont
  {Evangelista}(2010)}]{lenzi2010non}%
  \BibitemOpen
  \bibfield  {author} {\bibinfo {author} {\bibfnamefont {E.}~\bibnamefont
  {Lenzi}}, \bibinfo {author} {\bibfnamefont {C.}~\bibnamefont {Yednak}}, \
  and\ \bibinfo {author} {\bibfnamefont {L.}~\bibnamefont {Evangelista}},\
  }\bibfield  {title} {\enquote {\bibinfo {title} {Non-markovian diffusion and
  the adsorption-desorption process},}\ }\href@noop {} {\bibfield  {journal}
  {\bibinfo  {journal} {Physical Review E}\ }\textbf {\bibinfo {volume} {81}},\
  \bibinfo {pages} {011116} (\bibinfo {year} {2010})}\BibitemShut {NoStop}%
\bibitem [{\citenamefont {Mura}, \citenamefont {Taqqu},\ and\ \citenamefont
  {Mainardi}(2008)}]{mura2008non}%
  \BibitemOpen
  \bibfield  {author} {\bibinfo {author} {\bibfnamefont {A.}~\bibnamefont
  {Mura}}, \bibinfo {author} {\bibfnamefont {M.~S.}\ \bibnamefont {Taqqu}}, \
  and\ \bibinfo {author} {\bibfnamefont {F.}~\bibnamefont {Mainardi}},\
  }\bibfield  {title} {\enquote {\bibinfo {title} {Non-markovian diffusion
  equations and processes: analysis and simulations},}\ }\href@noop {}
  {\bibfield  {journal} {\bibinfo  {journal} {Physica A: Statistical Mechanics
  and its Applications}\ }\textbf {\bibinfo {volume} {387}},\ \bibinfo {pages}
  {5033--5064} (\bibinfo {year} {2008})}\BibitemShut {NoStop}%
\bibitem [{\citenamefont {Gao}\ \emph {et~al.}(2017)\citenamefont {Gao},
  \citenamefont {Wang}, \citenamefont {Shu}, \citenamefont {Gao},\ and\
  \citenamefont {Braunstein}}]{gao2017promoting}%
  \BibitemOpen
  \bibfield  {author} {\bibinfo {author} {\bibfnamefont {L.}~\bibnamefont
  {Gao}}, \bibinfo {author} {\bibfnamefont {W.}~\bibnamefont {Wang}}, \bibinfo
  {author} {\bibfnamefont {P.}~\bibnamefont {Shu}}, \bibinfo {author}
  {\bibfnamefont {H.}~\bibnamefont {Gao}}, \ and\ \bibinfo {author}
  {\bibfnamefont {L.~A.}\ \bibnamefont {Braunstein}},\ }\bibfield  {title}
  {\enquote {\bibinfo {title} {Promoting information spreading by using contact
  memory},}\ }\href@noop {} {\bibfield  {journal} {\bibinfo  {journal}
  {Europhysics Letters}\ }\textbf {\bibinfo {volume} {118}},\ \bibinfo {pages}
  {18001} (\bibinfo {year} {2017})}\BibitemShut {NoStop}%
\bibitem [{\citenamefont {Xue}\ and\ \citenamefont {Zhang}(2013)}]{xue2013non}%
  \BibitemOpen
  \bibfield  {author} {\bibinfo {author} {\bibfnamefont {P.}~\bibnamefont
  {Xue}}\ and\ \bibinfo {author} {\bibfnamefont {Y.-S.}\ \bibnamefont
  {Zhang}},\ }\bibfield  {title} {\enquote {\bibinfo {title} {Non-markovian
  decoherent quantum walks},}\ }\href@noop {} {\bibfield  {journal} {\bibinfo
  {journal} {Chinese Physics B}\ }\textbf {\bibinfo {volume} {22}},\ \bibinfo
  {pages} {070302} (\bibinfo {year} {2013})}\BibitemShut {NoStop}%
\bibitem [{\citenamefont {Casado-Pascual}\ \emph {et~al.}(2005)\citenamefont
  {Casado-Pascual}, \citenamefont {G{\'o}mez-Ord{\'o}{\~n}ez}, \citenamefont
  {Morillo}, \citenamefont {Lehmann}, \citenamefont {Goychuk},\ and\
  \citenamefont {H{\"a}nggi}}]{casado2005theory}%
  \BibitemOpen
  \bibfield  {author} {\bibinfo {author} {\bibfnamefont {J.}~\bibnamefont
  {Casado-Pascual}}, \bibinfo {author} {\bibfnamefont {J.}~\bibnamefont
  {G{\'o}mez-Ord{\'o}{\~n}ez}}, \bibinfo {author} {\bibfnamefont
  {M.}~\bibnamefont {Morillo}}, \bibinfo {author} {\bibfnamefont
  {J.}~\bibnamefont {Lehmann}}, \bibinfo {author} {\bibfnamefont
  {I.}~\bibnamefont {Goychuk}}, \ and\ \bibinfo {author} {\bibfnamefont
  {P.}~\bibnamefont {H{\"a}nggi}},\ }\bibfield  {title} {\enquote {\bibinfo
  {title} {Theory of frequency and phase synchronization in a rocked bistable
  stochastic system},}\ }\href@noop {} {\bibfield  {journal} {\bibinfo
  {journal} {Physical Review E}\ }\textbf {\bibinfo {volume} {71}},\ \bibinfo
  {pages} {011101} (\bibinfo {year} {2005})}\BibitemShut {NoStop}%
\bibitem [{\citenamefont {Lin}\ \emph {et~al.}(2020)\citenamefont {Lin},
  \citenamefont {Feng}, \citenamefont {Tang}, \citenamefont {Liu},
  \citenamefont {Xu}, \citenamefont {Hui},\ and\ \citenamefont
  {Lai}}]{lin2020non}%
  \BibitemOpen
  \bibfield  {author} {\bibinfo {author} {\bibfnamefont {Z.-H.}\ \bibnamefont
  {Lin}}, \bibinfo {author} {\bibfnamefont {M.}~\bibnamefont {Feng}}, \bibinfo
  {author} {\bibfnamefont {M.}~\bibnamefont {Tang}}, \bibinfo {author}
  {\bibfnamefont {Z.}~\bibnamefont {Liu}}, \bibinfo {author} {\bibfnamefont
  {C.}~\bibnamefont {Xu}}, \bibinfo {author} {\bibfnamefont {P.~M.}\
  \bibnamefont {Hui}}, \ and\ \bibinfo {author} {\bibfnamefont {Y.-C.}\
  \bibnamefont {Lai}},\ }\bibfield  {title} {\enquote {\bibinfo {title}
  {Non-markovian recovery makes complex networks more resilient against
  large-scale failures},}\ }\href@noop {} {\bibfield  {journal} {\bibinfo
  {journal} {Nature Communications}\ }\textbf {\bibinfo {volume} {11}},\
  \bibinfo {pages} {2490} (\bibinfo {year} {2020})}\BibitemShut {NoStop}%
\end{thebibliography}%

\end{document}